\documentclass[reprint,aps,prd,onecolumn,notitlepage,showpacs,nofootinbib,preprintnumbers,superscriptaddress]{revtex4-1}
\usepackage{amsmath}
\usepackage{amsfonts}
\usepackage{amssymb}
\usepackage{latexsym}
\usepackage{color,xcolor}
\usepackage{graphicx}
\usepackage{bm}
\usepackage{epsfig}
\usepackage[english]{babel}
\usepackage{hyperref}
\usepackage{times}
\usepackage{comment}

\def\O{\mathcal{O}}
\def\D{\mathcal{D}}
\def\GB{\mathrm{GB}}
\def\lagb{\lambda_{\GB}}
\def\Nsh{N_{\sharp}}
\def\ptr{(\pi TR)}
\def\pt{\pi T}
\def\fw{\mathfrak{w}}
\def\fq{\mathfrak{q}}
\def\fz{\mathfrak{z}}
\def\tom{\tilde{\omega}}
\def\tq{\tilde{q}}

\begin{document}

\title{Notes on diffusive and shear quasinormal modes of black branes}
\author{Yu-Chen Ding}
\author{Li Li}
\author{Towe Wang}
\email[Electronic address: ]{twang@phy.ecnu.edu.cn}
\affiliation{Department of Physics, East China Normal University,\\
Shanghai 200241, China\\ \vspace{0.2cm}}
\date{\today\\ \vspace{1cm}}
\begin{abstract}
In the literature, to extract the dispersion relation of low-frequency quasinormal modes in both diffusive and shear channels, it is a customary recipe to assume firstly $\omega\sim\mathcal{O}$ to solve the equation of motion and finally $\omega\sim\mathcal{O}(q^2)$ when applying the Dirichlet boundary condition. The two assumptions appear confusing though the recipe usually gives the same result as that from other channels or from the Kubo formula. We refine the recipe by assuming $\omega\sim\mathcal{O}(q^2)$ from the beginning to the end, and demonstrate it in the diffusive channel of the Schwarzschild black brane and the shear channel of the Gauss-Bonnet black brane.
\end{abstract}


\maketitle




\section{Introduction}\label{sect-intro}
The past twenty years have witnessed the exciting development of the AdS/CFT correspondence \cite{Maldacena:1997re,Aharony:1999ti}. Among many important results, a pragmatic one is the Kovtun-Starinets-Son bound \cite{Policastro:2001yc,Kovtun:2003wp,Buchel:2003tz,Kovtun:2004de} on $\eta/s$, the ratio of the shear viscosity to the entropy density. In the framework of the Minkowski AdS/CFT correspondence \cite{Son:2002sd}, there are several methods \cite{Policastro:2002se,Policastro:2002tn,Kovtun:2005ev,Son:2007vk} to extract the ratio $\eta/s$ by studying different channels of the electromagnetic and gravitational perturbations. In Ref. \cite{Kovtun:2005ev}, the spin-0 and spin-1 modes of the electromagnetic perturbations are termed as the diffusive (or longitudinal) and transverse channels respectively, while the spin-0, spin-1 and spin-2 modes of the gravitational perturbations are referred as the sound, shear and scalar channels accordingly. We will follow their jargon in our notes.

Here we are concerned with two concrete examples in the literature. One example is the diffusive quasinormal modes of the Schwarzschild black brane in the Einstein gravity \cite{Kovtun:2005ev,Starinets:2008fb}. The other example is the shear quasinormal modes of a black brane in the Gauss-Bonnet gravity \cite{Brigante:2007nu}. In both examples, the ratio $\eta/s$ in the dual field theory can be inferred from the dispersion relation of the low-frequency quasinormal modes. In Refs. \cite{Kovtun:2005ev,Starinets:2008fb,Brigante:2007nu}, to find analytical solutions in the low-frequency limit, the frequency $\omega$ and the momentum $q$ of the quasinormal modes are rescaled as $\omega\rightarrow\lambda\omega$, $q\rightarrow\lambda q$ with $\lambda\rightarrow0$, assuming firstly that $\omega$ and $q$ are of the same order, i.e. $\omega\sim\O(q)$. Finally, to get a diffusion coefficient or a damping constant of correct value, it was also assumed that $\omega$ and $q^2$ are of the same order therein, i.e. $\omega\sim\O(q^2)$. These assumptions are confusing in the low-frequency limit.

In our notes, we will rescale $\omega\rightarrow\lambda\omega$, $q^2\rightarrow\lambda q^2$ in accordance with $\omega\sim\O(q^2)$ and refine some details in Refs. \cite{Starinets:2008fb,Brigante:2007nu}. We will get rid of the assumption $\omega\sim\O(q)$ throughout the calculations. After more than a decade of development, the holographic calculation of $\eta/s$ has been well established. But it is never too late to refine the details. As we will see, by rescaling in the consistent manner, we can also eliminate some ambiguities in Refs. \cite{Starinets:2008fb,Brigante:2007nu}.


The outline of our notes is as follows. In the remainder of this section, as preparation, we will recall line elements of the Schwarzschild and the Gauss-Bonnet black branes as well as their Hawking temperature. In the diffusive channel, quasinormal modes of the Schwarzschild black brane will be studied in Sec. \ref{sect-Sch}. Treating the Gauss-Bonnet parameter $\lagb$ perturbatively in Sec. \ref{sect-GBsm}, we will investigate the quasinormal modes of the Gauss-Bonnet black brane in the shear channel. The calculations in Secs. \ref{sect-Sch}, \ref{sect-GBsm} are similar to Refs. \cite{Starinets:2008fb,Brigante:2007nu}, except for that we rescale the frequency and the momentum as $\omega\rightarrow\lambda\omega$, $q^2\rightarrow\lambda q^2$ throughout. For comparison, in Sec. \ref{sect-lin} we point out the ambiguities of some details in Refs. \cite{Starinets:2008fb,Brigante:2007nu} which took $\omega\rightarrow\lambda\omega$, $q\rightarrow\lambda q$ firstly and treated $\omega\sim\O(q^2)$ finally. The implications of the notes will be discussed in Sec. \ref{sect-disc} briefly. In Appendix \ref{app-lin}, we will explore the result of assuming $\fw\sim\O(\fq)$ without $\omega\sim\O(q^2)$.

In the ($4+1$)-dimensional spacetime, a nonrotating black brane is described by the line element
\begin{equation}\label{metricr}
ds^2=\frac{r^2}{R^2}\left[-\Nsh^2f(r)dt^2+dx^2+dy^2+dz^2\right]+\frac{R^2}{r^2f(r)}dr^2.
\end{equation}
For the Schwarzschild black brane in the Einstein gravity \cite{Kovtun:2005ev,Starinets:2008fb}, the function $f(r)$ and the parameter $\Nsh$ are given by
\begin{equation}\label{fSchr}
f(r)=1-\frac{r_0^4}{r^4},~~~~\Nsh^2=1,
\end{equation}
while for a black brane in the Gauss-Bonnet gravity \cite{Brigante:2007nu,Brigante:2008gz,Cai:2001dz,Konoplya:2017zwo,Konoplya:2017ymp,Konoplya:2017lhs} they are
\begin{equation}\label{fGBr}
f(r)=\frac{1}{2\lagb}\left[1-\sqrt{1-4\lagb\left(1-\frac{r_0^4}{r^4}\right)}\right],~~~~\Nsh^2=\frac{1}{2}\left(1+\sqrt{1-4\lagb}\right).
\end{equation}
Without loss of generality, we will take $\Nsh>0$. In both cases, the Horizon is located at $r=r_0$, and the Hawking temperature is of the form
\begin{equation}
T=\frac{\Nsh r_0}{\pi R^2}.
\end{equation}
It is apparent that the Gauss-Bonnet black brane reduces to the Schwarzschild black brane in the limit $\lagb\rightarrow0$. This limit will be taken in Sec. \ref{sect-GBsm} to check the consistency of our results. One can also take this limit to check the consistency between Sec. IVB2 of Ref. \cite{Kovtun:2005ev} and Appendix B1 of Ref. \cite{Brigante:2007nu}.

Throughout the notes, to save the space and notations, we will slightly abuse the notation of Lebesgue-Stieltjes integration. For example, in our conventions,
\begin{equation}
\int_{\infty}^{\fz}f(\fz)dg(\fz)\doteq\int_{g(\infty)}^{g(\fz)}f(\fz')dg(\fz')=\int_{\infty}^{\fz}f(\fz')\frac{dg(\fz')}{d\fz'}d\fz'.
\end{equation}
Focusing always on the small $\lagb$ case, we will use $\approx$ to denote equivalence neglecting $\O(\lagb^2)$ terms.

\section{Diffusive channel of Schwarzschild black brane}\label{sect-Sch}
For comparison with Refs. \cite{Kovtun:2005ev,Starinets:2008fb}, this section will follow closely the notation and treatment in Ref. \cite{Starinets:2008fb}. However, at the beginning, we will keep $\Nsh$ explicit in relation to Sec. \ref{sect-GBsm}.

For the Schwarzschild black brane, we introduce a dimensionless coordinate $u=r_0^2/r^2$, with respect to which the derivative will be denoted by prime. In terms of $u$, Eqs. \eqref{metricr} and \eqref{fSchr} can be rewritten as
\begin{eqnarray}
&&ds^2=\frac{\ptr^2}{u}\left[-f(u)dt^2+\frac{dx^2+dy^2+dz^2}{\Nsh^2}\right]+\frac{R^2}{4u^2f(u)}du^2,\label{metricu}\\
&&f(u)=1-u^2,~~~~\Nsh^2=1.\label{fSchu}
\end{eqnarray}
It is useful in this section to write down the following combinations
\begin{eqnarray}\label{comb}
\nonumber&&\sqrt{-g}=\frac{R\ptr^4}{2\Nsh^3u^3},~~~~,\frac{g^{tt}}{g^{xx}}=-\frac{1}{\Nsh^2f},~~~~\frac{g^{xx}}{g^{uu}}=\frac{\Nsh^2}{(2\pt)^2uf},\\
&&\sqrt{-g}g^{tt}g^{uu}=-\frac{2\ptr^2}{\Nsh^3R},~~~~\sqrt{-g}g^{tt}g^{xx}=-\frac{R}{2\Nsh uf}.
\end{eqnarray}

The electric field  parallel to the brane, $E_x=\omega A_x+qA_t$, proves to satisfy the equation of motion
\begin{equation}
E_x''+\partial_u\left[\ln\left(\sqrt{-g}g^{tt}g^{uu}\right)-\ln\left(\frac{\omega^2g^{tt}}{g^{xx}}+q^2\right)\right]E_x'-\frac{\omega^2g^{tt}+q^2g^{xx}}{g^{uu}}E_x=0,
\end{equation}
which can be written in the form \cite{Starinets:2008fb}
\begin{equation}\label{eomSch}
E_x''+\left[\frac{\fw^2f'}{f(\fw^2-\Nsh^2\fq^2f)}+\partial_u\ln\left(\sqrt{-g}g^{tt}g^{uu}\right)\right]E_x'+\frac{(2\pt)^2g^{xx}}{fg^{uu}}\left(\Nsh^{-2}\fw^2-\fq^2f\right)E_x=0
\end{equation}
where
\begin{equation}\label{fwfq}
\fw=\frac{\omega}{2\pt}=\frac{R^2\omega}{2\Nsh r_0},~~~~\fq=\frac{q}{2\pt}=\frac{R^2q}{2\Nsh r_0}.
\end{equation}

Near the horizon, $u\rightarrow1$, $f(u)\rightarrow0$, one can check that $f'(u)\rightarrow-2$, and that the dominant terms in Eq. \eqref{eomSch} are
\begin{equation}
E_x''+\frac{f'}{f}E_x'+\frac{\fw^2}{f^2}E_x=0.
\end{equation}
At the horizon, we impose the boundary condition that only infalling waves survive. In accordance with this condition, the solution to Eq. \eqref{eomSch} should take the form \cite{Starinets:2008fb}
\begin{equation}\label{decE}
E_x(u)=f^{-i\fw/2}F(u)
\end{equation}
with $F(u)$ being regular at $u=1$. At spatial infinity, $u=0$, the Dirichlet boundary condition $E_x(0)=0$ should be imposed \cite{Starinets:2008fb}.\footnote{This is not what one has to do. One first sets the boundary condition $E_x(0)=C$, where $C$ is a function of $\omega$ and $q$. Then $C=0$ is the additional condition for finding quasinormal modes on top of the Dirichlet condition. We thank the anonymous referee for bringing this fact to our attention. In Secs. \ref{sect-Sch}, \ref{sect-GBsm}, this fact can be confirmed by studying the divergent terms in the limit $u\rightarrow1$ with the coefficient $C_0$.} According to Eqs. \eqref{fSchu}, we will set $\Nsh=1$ in the rest of this section.

All we have presented so far are the same as in Ref. \cite{Starinets:2008fb}. To proceed, for the reason given in Sec. \ref{sect-intro}, we will study the hydrodynamic limit
by assuming $\fw\sim\O(\fq^2)$ throughout this section. In contrast, Refs. \cite{Kovtun:2005ev,Starinets:2008fb} assumed firstly $\fw\sim\O(\fq)$ and finally $\fw\sim\O(\fq^2)$, see Sec. IVA of Ref. \cite{Kovtun:2005ev} as well as Sec. 2 of Ref. \cite{Starinets:2008fb}, which have ambiguities in some details as we will examine in Sec. \ref{subsect-Star} honestly.

In the low-energy limit $\omega/T\rightarrow0$, $q/T\rightarrow0$, we can formally expand the dispersion relation of quasinormal modes as
\begin{equation}\label{dispexp}
\omega=C^{(1)}q+C^{(2)}q^2+\cdots.
\end{equation}
We will set $C^{(1)}=0$ in the current and the next sections. Otherwise, as shown in Appendix \ref{app-lin}, a nonvanishing $O(q)$ term in the dispersion relation leads to vanishing wave functions up to $O(q)$ at the least.

To trace the order of $\fq$, it is convenient to introduce a book-keeping parameter $\lambda$ as in Ref. \cite{Starinets:2008fb} and rescale $\fq^2\rightarrow\lambda\fq^2$. As we have set $C^{(1)}=0$, the hydrodynamic limit corresponds to the $O(q^2)$ term in Eq. \eqref{dispexp}, which can be captured by rescaling the frequency as $\fw\rightarrow\lambda\fw$ at the same time. After rescaling and expanding in powers of $\lambda$, we find the solution \eqref{decE} turns out to be
\begin{equation}
E_x(u,\lambda\fw,\lambda\fq^2)=\left(1-\frac{i\lambda\fw}{2}\ln f\right)\left[F_0(u)+\lambda F_1(u)\right]+\O(\lambda^2),
\end{equation}
and the equation of motion \eqref{eomSch} becomes
\begin{eqnarray}
\nonumber&&F_0''+\left[\partial_u\ln\left(\sqrt{-g}g^{tt}g^{uu}\right)\right]F_0'\\
\nonumber&&+\lambda\left\{\partial_u^2\left[F_1-\left(\frac{i\fw}{2}\ln f\right)F_0\right]+\left[\partial_u\ln\left(\sqrt{-g}g^{tt}g^{uu}\right)\right]\partial_u\left[F_1-\left(\frac{i\fw}{2}\ln f\right)F_0\right]-\frac{\fw^2f'}{\fq^2f^2}F_0'-\frac{(2\pt)^2\fq^2g^{xx}}{g^{uu}}F_0\right\}\\
&&+\O(\lambda^2)=0.
\end{eqnarray}
By definition functions $F_0(u)$ and $F_1(u)$ are free of $\lambda$, and the equation of motion is expected to hold in each order of $\lambda$. Hence this equation can be solved order by order, yielding
\begin{eqnarray}
F_0&=&C_1\int_0^u\frac{1}{\sqrt{-g}g^{tt}g^{uu}}du+C_0,\label{F0}\\
\nonumber F_1-\left(\frac{i\fw}{2}\ln f\right)F_0&=&\int_0^u\left(\frac{1}{\sqrt{-g}g^{tt}g^{uu}}\int_0^u\frac{\fw^2f'}{\fq^2f^2}C_1du\right)du+C_3\int_0^u\frac{1}{\sqrt{-g}g^{tt}g^{uu}}du+C_2\\
&&+\int_0^u\left\{\frac{1}{\sqrt{-g}g^{tt}g^{uu}}\int_0^u\left[(2\pt)^2\fq^2\sqrt{-g}g^{tt}g^{xx}\left(C_1\int_0^u\frac{1}{\sqrt{-g}g^{tt}g^{uu}}du+C_0\right)\right]du\right\}du.\label{F1}
\end{eqnarray}
Note that $\sqrt{-g}g^{tt}g^{uu}$ is a constant as shown in Eqs. \eqref{comb}.

In the above solution, $C_0$, $C_1$, $C_2$, $C_3$ are constants of integration. Some of them can be fixed by boundary conditions. The boundary conditions are expected to be satisfied in each order of $\lambda$. Setting the upper limit of integration to 0, it reads that
\begin{equation}
E_x(0,\lambda\fw,\lambda\fq^2)=C_0+\lambda C_2+\O(\lambda^2).
\end{equation}
Consequently, the Dirichlet boundary condition $E_x(0)=0$ fixes the values $C_0=C_2=0$. In the limit $u\rightarrow1$, $F(u)$ should be regular according to the boundary condition. Therefore, both $F_0(u)$ and $F_1(u)$ are expected to be regular at $u=1$. It is trivial to check that $F_0(1)$ is finite. However, the finiteness of $F_1(1)$ is nontrivial.

As we will clarify at the end of this section, the last line of Eq. \eqref{F1} is finite in the limit $u\rightarrow1$. Since $\sqrt{-g}g^{tt}g^{uu}$ is a constant, the term proportional to $C_3$ in Eq. \eqref{F1} is apparently finite at $u=1$. So $F_1(1)$ is finite if the divergent part in
\begin{equation}
\left[\frac{i\fw}{2}\ln f(1)\right]C_1\int_0^1\frac{1}{\sqrt{-g}g^{tt}g^{uu}}du+\int_0^1\left(\frac{1}{\sqrt{-g}g^{tt}g^{uu}}\int_0^u\frac{\fw^2f'}{\fq^2f^2}C_1du\right)du
\end{equation}
is cancelled. This expression can be reformed as
\begin{equation}
\left(\frac{i\fw}{2}\int_0^1\frac{f'}{f}du\right)C_1\int_0^1\frac{1}{\sqrt{-g}g^{tt}g^{uu}}du+\int_0^1\left[\frac{1}{\sqrt{-g}g^{tt}g^{uu}}\frac{\fw^2}{\fq^2}C_1\left(1-\frac{1}{f}\right)\right]du.
\end{equation}
Clearly, divergence is induced simply by the factor $1/f$ in the integrand at $u=1$. Thus the cancellation of divergence requires
\begin{equation}
\frac{i\fw f'(1)}{2}\int_0^1\frac{C_1}{\sqrt{-g}g^{tt}g^{uu}}du-\frac{1}{\sqrt{-g}g^{tt}g^{uu}}\frac{\fw^2C_1}{\fq^2}=0.
\end{equation}
Rewritten in terms of $\omega$ and $q$ via Eqs. \eqref{fwfq}, it means the $O(q^2)$ term in Eq. \eqref{dispexp} is given by
\begin{eqnarray}
\nonumber C^{(2)}&=&-i\frac{1}{2\pt}\sqrt{-g}g^{tt}g^{uu}\int_0^1\frac{1}{\sqrt{-g}g^{tt}g^{uu}}du\\
&=&-i\frac{g^{xx}(1)}{\sqrt{-g^{tt}(1)g^{uu}(1)}}\sqrt{-g(1)}g^{tt}(1)g^{uu}(1)\int_0^1\frac{1}{\sqrt{-g}g^{tt}g^{uu}}du.
\end{eqnarray}
In the second line, Eqs. \eqref{comb} have been utilized. Indeed, in the diffusive channel \cite{Starinets:2008fb}, the lowest quasinormal frequency behaves as $\omega=-iDq^2+\O(q^4)$, where the diffusion constant is
\begin{equation}\label{DifSch}
D=\frac{\sqrt{-g(1)}}{g_{xx}(1)\sqrt{-g_{tt}(1)g_{uu}(1)}}\int_0^1\frac{-g_{tt}(u)g_{uu}(u)}{\sqrt{-g(u)}}du.
\end{equation}
In the Einstein gravity, it can be translated into the damping constant in the shear channel \cite{Kovtun:2003wp,Son:2007vk,Starinets:2008fb}
\begin{equation}\label{DamSch}
\D=\frac{\sqrt{-g(1)}}{\sqrt{-g_{tt}(1)g_{uu}(1)}}\int_0^1\frac{-g_{tt}(u)g_{uu}(u)}{g_{xx}(u)\sqrt{-g(u)}}du.
\end{equation}
For the Schwarzschild black brane Eq. \eqref{metricu}, this implies the ratio $\eta/s=1/(4\pi)$ of the shear viscosity to the entropy density in the dual theory \cite{Son:2002sd,Kovtun:2005ev}.

In the above, we have used the fact that the last line of Eq. \eqref{F1} is finite in the limit $u\rightarrow1$. To verify this fact, we set $C_0=0$ as dictated by the Dirichlet boundary condition $E_x(0)=0$, and then substitute Eqs. \eqref{comb} into the last line of \eqref{F1},
\begin{eqnarray}\label{finF}
\nonumber-\frac{C_1\Nsh^5R\fq^2}{2\ptr^2}\int_0^1\left\{\int_0^u\left[\frac{1}{uf}\left(\int_0^udu\right)\right]du\right\}du&=&-\frac{2C_1\Nsh^5\fq^2}{(2\pt)^2R}\int_0^1\left(\int_0^u\frac{1}{f}du\right)du\\
\nonumber&=&-\frac{C_1\Nsh^5R\fq^2}{2\ptr^2}\int_0^1\frac{1}{f}du+\frac{C_1\Nsh^5R\fq^2}{2\ptr^2}\int_0^1\frac{u}{f}du\\
&=&-\frac{C_1\Nsh^5R\fq^2}{2\ptr^2}\int_0^1\frac{1}{1+u}du
\end{eqnarray}
In the second step, we have made the integration by parts. In the last step, Eqs. \eqref{fSchu} have been inserted. It is clear that the integral Eq. \eqref{finF} is not divergent. In other words, the last line of Eq. \eqref{F1} is finite at $u=1$ as promised.

\section{Shear channel of Gauss-Bonnet black brane: small $\lagb$}\label{sect-GBsm}
In accordance with Ref. \cite{Brigante:2007nu}, when studying the Gauss-Bonnet black brane, we will work with the notations
\begin{equation}
\fz=\frac{r}{r_0},~~~~\tom=\frac{R^2\omega}{r_0}=\frac{\Nsh\omega}{\pt},~~~~\tq=\frac{R^2q}{r_0}=\frac{\Nsh q}{\pt}.
\end{equation}
Then Eqs. \eqref{metricr}, \eqref{fGBr} turn out to be
\begin{eqnarray}
&&ds^2=\ptr^2\fz^2\left[-f(\fz)dt^2+\frac{dx^2+dy^2+dz^2}{\Nsh^2}\right]+\frac{R^2}{\fz^2f(\fz)}d\fz^2,\label{metricz}\\
&&f(\fz)=\frac{1}{2\lagb}\left[1-\sqrt{1-4\lagb\left(1-\frac{1}{\fz^4}\right)}\right],~~~~\Nsh^2=\frac{1}{2}\left(1+\sqrt{1-4\lagb}\right).\label{fGBz}
\end{eqnarray}
In our notes, we follow closely the notations in Ref. \cite{Starinets:2008fb}. The notations in Ref. \cite{Brigante:2007nu} are slightly different from those in Ref. \cite{Starinets:2008fb}. To be self-consistent, we have to change some notations and conventions in Ref. \cite{Brigante:2007nu}, although we keep as more as possible. When making comparisons, one should notice that $f(r)$ in Ref. \cite{Brigante:2007nu} is replaced by $r^2f(r)/R^2$ in our notes. What is more, $x_1,x_2,x_3$, $L$, $z$ are replaced by $y,z,x$, $R$, $\fz$ correspondingly.

In the Gauss-Bonnet gravity, the shear channel equations can be reduced to a single equation for $Z=qg^{yy}h_{ty}+\omega g^{yy}h_{xy}$. At first order in $\lagb$, it is \cite{Brigante:2007nu}
\begin{eqnarray}\label{eomGBsm}
\nonumber&&\partial_{\fz}^2Z+\left[\frac{5\fz^4-1}{\fz^4-1}+\frac{4\tq^2}{\tq^2(-\fz^4+1)+\Nsh^{-2}\tom^2\fz^4}\right]\frac{1}{\fz}\partial_{\fz}Z+\frac{\tq^2(-\fz^4+1)+\Nsh^{-2}\tom^2\fz^4}{(\fz^4-1)^2}Z\\
&&+4\lagb\left\{-\frac{2\tq^4(\fz^4-1)^2+4\Nsh^{-2}\tom^2\tq^2\fz^4-3\Nsh^{-4}\tom^4\fz^8}{\fz^5\left[\tq^2(\fz^4-1)-\Nsh^{-2}\tom^2\fz^4\right]^2}\partial_{\fz}Z+\frac{\tq^2(\fz^4+3)-2\Nsh^{-2}\tom^2\fz^4}{4\fz^4(\fz^4-1)}Z\right\}\approx0.
\end{eqnarray}
Keep in mind that $\approx$ denotes equivalence neglecting $\O(\lagb^2)$ terms. Our notes focus on the small $\lagb$ case for two reasons. First, this case was studied in Appendix B1 of Ref. \cite{Brigante:2007nu}, thus a direct comparison is possible. Second, this enables a calculation perturbative for $\lagb$, which is much simpler than a nonperturbative calculation. The nonperturbative calculation for arbitrary $\lagb$ is very tedious and thus presented in an accompanying paper \cite{Wang:2019xny}.

We will study the hydrodynamic limit by assuming $\tom\sim\O(\tq^2)$ in the whole section. In contrast, Ref. \cite{Brigante:2007nu} assumed firstly $\tom\sim\O(\tq)$ and finally $\tom\sim\O(\tq^2)$, see Appendix B1 of Ref. \cite{Brigante:2007nu}, which has ambiguities as we will point out in Sec. \ref{subsect-Liu}.

Near the horizon, $\fz\rightarrow1$, the divergent terms in Eq. \eqref{eomGBsm} are dominated by
\begin{equation}
\partial_{\fz}^2Z+\frac{4\fz^4}{\fz^4-1}\partial_{\fz}Z+\frac{\Nsh^{-2}\tom^2\fz^8}{(\fz^4-1)^2}Z\approx0.
\end{equation}
At the horizon, again we impose the boundary condition that only infalling waves survive. Accordingly, the solution to Eq. \eqref{eomGBsm} can be put in the form
\begin{equation}\label{decZsm}
Z(\fz)\approx\left(1-\frac{1}{\fz^4}\right)^{-i\tom/(4\Nsh)}g(\fz)
\end{equation}
with $g(\fz)$ being regular at $\fz=1$. At spatial infinity, $\fz\rightarrow\infty$, we impose the Dirichlet boundary condition $Z(\infty)=0$.

Parallel to the previous section, after rescaling $\tom\rightarrow\lambda\tom$, $\tq^2\rightarrow\lambda\tq^2$ and expanding in powers of $\lambda$, we rewrite the solution \eqref{decZsm} as
\begin{equation}
Z(\fz,\lambda\tom,\lambda\tq^2)\approx\left(1-\frac{i\lambda\tom}{4\Nsh}\ln\frac{\fz^4-1}{\fz^4}\right)\left[g_0(\fz)+\lambda g_1(\fz)\right]+\O(\lambda^2)
\end{equation}
and the equation of motion \eqref{eomGBsm} as
\begin{eqnarray}
\nonumber&&\partial_{\fz}^2g_0+\frac{5}{\fz}\partial_{\fz}g_0-\frac{8\lagb}{\fz^5}\partial_{\fz}g_0\\
\nonumber&&+\lambda\Biggl\{\partial_{\fz}^2\left[g_1-\left(\frac{i\tom}{4\Nsh}\ln\frac{\fz^4-1}{\fz^4}\right)g_0\right]+\frac{5}{\fz}\partial_{\fz}\left[g_1-\left(\frac{i\tom}{4\Nsh}\ln\frac{\fz^4-1}{\fz^4}\right)g_0\right]-\frac{4\tom^2\fz^3}{\Nsh^2\tq^2(\fz^4-1)^2}\partial_{\fz}g_0-\frac{\tq^2}{\fz^4-1}g_0\\
\nonumber&&-\frac{8\lagb}{\fz^5}\partial_{\fz}\left[g_1-\left(\frac{i\tom}{4\Nsh}\ln\frac{\fz^4-1}{\fz^4}\right)g_0\right]+\lagb\left[-\frac{16\tom^2\fz^3}{\Nsh^2\tq^2(\fz^4-1)^2}\partial_{\fz}g_0+\frac{\tq^2(\fz^4+3)}{\fz^4(\fz^4-1)}g_0\right]\Biggr\}\\
&&+\O(\lambda^2)\approx0.
\end{eqnarray}
Order by order, the equation of motion is separated into
\begin{eqnarray}
\nonumber&&\frac{1}{\fz^5e^{2\lagb/\fz^4}}\partial_{\fz}\left(\fz^5e^{2\lagb/\fz^4}\partial_{\fz}g_0\right)\approx0,\\
\nonumber&&\frac{1}{\fz^5e^{2\lagb/\fz^4}}\partial_{\fz}\left\{\fz^5e^{2\lagb/\fz^4}\partial_{\fz}\left[g_1-\left(\frac{i\tom}{4\Nsh}\ln \frac{\fz^4-1}{\fz^4}\right)g_0\right]\right\}-\frac{4\tom^2\fz^3(1+4\lagb)}{\Nsh^2\tq^2(\fz^4-1)^2}\partial_{\fz}g_0+\frac{\tq^2\left[-\fz^4+\lagb(\fz^4+3)\right]}{\fz^4(\fz^4-1)}g_0\approx0\\
\end{eqnarray}
and solved by
\begin{eqnarray}
g_0&\approx&\frac{C_1}{2\lagb}\left(e^{-2\lagb/\fz^4}-1\right)+C_0\approx C_1\left(-\frac{1}{\fz^4}+\frac{\lagb}{\fz^8}\right)+C_0,\label{g0sm}\\
\nonumber g_1-\left(\frac{i\tom}{4\Nsh}\ln \frac{\fz^4-1}{\fz^4}\right)g_0&\approx&\int_{\infty}^{\fz}\left[\frac{1}{\fz(\fz^4+2\lagb)}\int_{\infty}^{\fz}\frac{16C_1\tom^2\fz^3}{\Nsh^2\tq^2(\fz^4-1)^2}(1+4\lagb)d\fz\right]d\fz+C_3\left(-\frac{1}{\fz^4}+\frac{\lagb}{\fz^8}\right)+C_2\\
&&+\int_{\infty}^{\fz}\left\{\frac{1}{\fz(\fz^4+2\lagb)}\int_{\infty}^{\fz}\frac{\tq^2\fz\left[\fz^4-\lagb(\fz^4+1)\right]}{\fz^4-1}\left[C_1\left(-\frac{1}{\fz^4}+\frac{\lagb}{\fz^8}\right)+C_0\right]d\fz\right\}d\fz.\label{g1sm}
\end{eqnarray}

Setting the upper limit of integration to $\infty$,
\begin{equation}
Z(\infty,\tom,\tq)\approx C_0+\lambda C_2+\O(\lambda^2),
\end{equation}
we can use the Dirichlet boundary condition $Z(\infty)=0$ to fix $C_0=C_2=0$.

Similar to the previous section, the regularity of $g(\fz)$ at $\fz=1$ demands the finiteness of both $g_0(1)$ and $g_1(1)$. The finiteness of $g_0(1)$ is obvious. To study the finiteness of $g_1(1)$, we take the limit $\fz\rightarrow1$ of Eq. \eqref{g1sm}. In this limit, the term proportional to $C_3$ is apparently finite, and the last line is also finite as to be verified at the end of this section. Then the divergent terms in $g_1(1)$ are
\begin{equation}
\left.\left(\frac{i\tom}{4\Nsh}\ln \frac{\fz^4-1}{\fz^4}\right)C_1\left(-\frac{1}{\fz^4}+\frac{\lagb}{\fz^8}\right)\right|_{\fz=1}+\int_{\infty}^1\left[\frac{1}{\fz(\fz^4+2\lagb)}\int_{\infty}^{\fz}\frac{16C_1\tom^2\fz^3}{\Nsh^2\tq^2(\fz^4-1)^2}(1+4\lagb)d\fz\right]d\fz.
\end{equation}
Note that the second term can be simplified by integration by parts,
\begin{eqnarray}\label{divgs}
\nonumber&&\int_{\infty}^1\left[\frac{1}{\fz(\fz^4+2\lagb)}\int_{\infty}^{\fz}\frac{16C_1\tom^2\fz^3}{\Nsh^2\tq^2(\fz^4-1)^2}(1+4\lagb)d\fz\right]d\fz\\
\nonumber&=&\frac{1}{2\lagb}\int_{\infty}^1\left[\int_{\infty}^{\fz}\frac{4C_1\tom^2\fz^3}{\Nsh^2\tq^2(\fz^4-1)^2}(1+4\lagb)d\fz\right]d\ln\frac{\fz^4}{\fz^4+2\lagb}\\
\nonumber&=&\frac{1}{2\lagb}\left(\ln\frac{1}{1+2\lagb}\right)\int_{\infty}^1\frac{4C_1\tom^2\fz^3}{\Nsh^2\tq^2(\fz^4-1)^2}(1+4\lagb)d\fz\\
\nonumber&&-\frac{1}{2\lagb}\int_{\infty}^1\left(\ln\frac{\fz^4}{\fz^4+2\lagb}\right)\frac{4C_1\tom^2\fz^3}{\Nsh^2\tq^2(\fz^4-1)^2}(1+4\lagb)d\fz\\
&\approx&-\int_{\infty}^1\frac{4C_1\tom^2\fz^3}{\Nsh^2\tq^2(\fz^4-1)}\left[1+\lagb\left(3-\frac{1}{\fz^4}\right)\right]d\fz.
\end{eqnarray}
To cancel the divergent terms in $g_1(1)$,
\begin{equation}
\left.\frac{i\tom}{4\Nsh}\frac{4\fz^3}{\fz^4-1}C_1\left(-\frac{1}{\fz^4}+\frac{\lagb}{\fz^8}\right)\right|_{\fz=1}-\left.\frac{4C_1\tom^2\fz^3}{\Nsh^2\tq^2(\fz^4-1)}\left[1+\lagb\left(3-\frac{1}{\fz^4}\right)\right]\right|_{\fz=1}
\end{equation}
should be finite. Barring the trivial case $C_1=0$, this requirement is equivalent to
\begin{equation}
i(-1+\lagb)-\frac{4}{\Nsh}(1+2\lagb)\frac{\pt\omega}{\Nsh q^2}=0.
\end{equation}
Therefore, similar to Eq. \eqref{dispexp} in the diffusive channel, the dispersion relation in the shear channel \cite{Kovtun:2003wp,Son:2007vk,Starinets:2008fb} is
\begin{equation}
\omega\approx-iq^2\frac{\Nsh^2(1-3\lagb)}{4\pt}+\O(q^3)
\end{equation}
which gives directly the damping constant
\begin{equation}\label{DamGBsm}
\D\approx\frac{\Nsh^2(1-3\lagb)}{4\pt}.
\end{equation}
The value of $\Nsh^2$ is given by Eqs. \eqref{fGBz}. As a result, the ratio of the shear viscosity to the entropy density is $\eta/s\approx(1-4\lagb)/(4\pi)$ in the dual theory of the Gauss-Bonnet black brane \cite{Brigante:2007nu,Brigante:2008gz}. As a consistency test, in the limit $\lagb\rightarrow0$, one can reproduce the damping constant $\D$ and the ratio $\eta/s$ in Sec. \ref{sect-Sch}.

To fulfill this section, now let us confirm the finiteness of the last line of Eq. \eqref{g1sm} at $\fz=1$. We set $C_0=0$, and then make the integration by parts in the same manner as Eq. \eqref{divgs},
\begin{eqnarray}\label{fings}
\nonumber&&\frac{C_1}{8\lagb}\int_{\infty}^1\left\{\left(\frac{1}{\fz^4}-\frac{1}{\fz^4+2\lagb}\right)\int_{\infty}^{\fz}\frac{\tq^2\fz\left[\fz^4-\lagb(\fz^4+1)\right]}{\fz^4-1}\left(\frac{1}{\fz^4}+\frac{\lagb}{\fz^8}\right)d\fz\right\}d\fz^4\\
\nonumber&=&\frac{C_1}{8\lagb}\int_{\infty}^1\left\{\int_{\infty}^{\fz}\frac{\tq^2\fz\left[\fz^4-\lagb(\fz^4+1)\right]}{\fz^4-1}\left(\frac{1}{\fz^4}+\frac{\lagb}{\fz^8}\right)d\fz\right\}d\ln\frac{\fz^4}{\fz^4+2\lagb}\\
\nonumber&=&\frac{C_1}{8\lagb}\left(\ln\frac{1}{1+2\lagb}\right)\int_{\infty}^1\frac{\tq^2\fz\left[\fz^4-\lagb(\fz^4+1)\right]}{\fz^4-1}\left(\frac{1}{\fz^4}+\frac{\lagb}{\fz^8}\right)d\fz\\
\nonumber&&-\frac{C_1}{8\lagb}\int_{\infty}^1\left(\ln\frac{\fz^4}{\fz^4+2\lagb}\right)\frac{\tq^2\fz\left[\fz^4-\lagb(\fz^4+1)\right]}{\fz^4-1}\left(\frac{1}{\fz^4}+\frac{\lagb}{\fz^8}\right)d\fz\\
&\approx&-\frac{C_1}{4}\int_{\infty}^1\frac{\tq^2}{\fz^3}\left[1-\lagb\left(2+\frac{1}{\fz^4}\right)\right]d\fz.
\end{eqnarray}
As can be seen from the final result, the integral Eq. \eqref{fings} is not divergent. In other words, the last line of Eq. \eqref{g1sm} is finite at $\fz=1$ as expected.

\section{Hydrodynamic limit assuming firstly $\fw\sim\O(\fq)$ and finally $\fw\sim\O(\fq^2)$}\label{sect-lin}
\subsection{Diffusive channel}\label{subsect-Star}
In Refs. \cite{Kovtun:2005ev,Starinets:2008fb}, the diffusion constant \eqref{DifSch} was verified by assuming firstly $\fw\sim\O(\fq)$ and finally $\fw\sim\O(\fq^2)$. There are some ambiguities in Sec. IVA of Ref. \cite{Kovtun:2005ev} and Sec. 2 of Ref. \cite{Starinets:2008fb}. In this subsection, we will examine these details. Before starting, we point out that the first term of Eq. (20) in Ref. \cite{Starinets:2008fb} should be $-\fw$, otherwise the derived diffusion constant will have a wrong sign.

Firstly, in accordance with the assumption $\fw\sim\O(\fq)$, we rescale the frequency and the momentum as $\fw\rightarrow\lambda\fw$, $\fq\rightarrow\lambda\fq$. Expanded in powers of $\lambda$, the solution \eqref{decE} and the equation of motion \eqref{eomSch} become \cite{Starinets:2008fb}
\begin{equation}
E_x(u,\lambda\fw,\lambda\fq)=\left(1-\frac{i\lambda\fw}{2}\ln f\right)\left[F_0(u)+\lambda F_1(u)\right]+\O(\lambda^2),
\end{equation}
\begin{eqnarray}
\nonumber&&F_0''+\left[\frac{\fw^2f'}{f(\fw^2-\fq^2f)}+\partial_u\ln\left(\sqrt{-g}g^{tt}g^{uu}\right)\right]F_0'\\
\nonumber&&+\lambda\left\{\partial_u^2\left[F_1-\left(\frac{i\fw}{2}\ln f\right)F_0\right]+\left[\frac{\fw^2f'}{f(\fw^2-\fq^2f)}+\partial_u\ln\left(\sqrt{-g}g^{tt}g^{uu}\right)\right]\partial_u\left[F_1-\left(\frac{i\fw}{2}\ln f\right)F_0\right]\right\}\\
&&+\O(\lambda^2)=0.
\end{eqnarray}
Note that the $\O(1)$ and $\O(\lambda)$ terms in this equation are different from their counterparts in Sec. \ref{sect-Sch}. In this subsection, we set $\Nsh=1$ in accordance with Eqs. \eqref{fSchu} and Ref. \cite{Starinets:2008fb}. Solving the above equation order by order, we obtain
\begin{eqnarray}
F_0&=&C_1\int_0^u\frac{\fw^2-\fq^2f}{f\sqrt{-g}g^{tt}g^{uu}}du+C_0,\\
F_1-\left(\frac{i\fw}{2}\ln f\right)F_0&=&C_3\int_0^u\frac{\fw^2-\fq^2f}{f\sqrt{-g}g^{tt}g^{uu}}du+C_2.
\end{eqnarray}

In Ref. \cite{Starinets:2008fb}, $F_0(1)$ and $F_1(1)$ are made regular by setting $C_1=0$ and requiring that
\begin{equation}
\frac{i\fw C_0}{2}\ln f(1)+C_3\int_0^1\frac{\fw^2-\fq^2f}{f\sqrt{-g}g^{tt}g^{uu}}du
\end{equation}
is finite. That is to say, the divergent part is cancelled,
\begin{equation}
\frac{i\fw}{2}f'(1)+\frac{C_3}{C_0}\frac{\fw^2}{\sqrt{-g}g^{tt}g^{uu}}=0.
\end{equation}
Eliminating $C_1$ and $C_3$ with these regularity conditions, we find
\begin{eqnarray}
F_0&=&C_0,\label{F0Star}\\
F_1&=&\frac{i\fw C_0}{2}\ln f-\frac{i}{2\fw}f'(1)C_0\sqrt{-g}g^{tt}g^{uu}\int_0^u\frac{\fw^2-\fq^2f}{f\sqrt{-g}g^{tt}g^{uu}}du+C_2,\label{F1Star}
\end{eqnarray}
and consequently
\begin{equation}\label{ExStar}
E_x(u,\fw,\fq)=C_0-\frac{i}{2\fw}f'(1)C_0\sqrt{-g}g^{tt}g^{uu}\int_0^u\frac{\fw^2-\fq^2f}{f\sqrt{-g}g^{tt}g^{uu}}du+C_2+\O(\fq^2).
\end{equation}

As claimed around Eq. (20) in Ref. \cite{Starinets:2008fb}, it is expected that the Dirichlet boundary condition $E_x(0)=0$ dictates the dispersion relation
\begin{equation}\label{DStar}
-1+\frac{i\fq^2}{2\fw}f'(1)\sqrt{-g}g^{tt}g^{uu}\int_0^1\frac{1}{\sqrt{-g}g^{tt}g^{uu}}du+\O(\fw)=0
\end{equation}
where we have corrected the sign as explained above and need to assume $\fw\sim\O(\fq^2)$ now. However, what we can really get from $E_x(0)=0$ is
\begin{equation}
C_0+C_2+\O(\fq^2)=0,
\end{equation}
which can be substituted into Eq. \eqref{F1Star} to yield
\begin{equation}
F_1=\frac{i\fq^2}{2\fw}f'(1)C_0\sqrt{-g}g^{tt}g^{uu}\int_0^u\frac{1}{\sqrt{-g}g^{tt}g^{uu}}du-C_0+\O(\fw).
\end{equation}
In view of this expression, to get Eq. \eqref{DStar}, one may further require that $F_1(1)=0$. As we will discuss in the coming subsection, a similar requirement was imposed implicitly in Ref. \cite{Brigante:2007nu} in the shear channel of Gauss-Bonnet black brane.

\subsection{Shear channel}\label{subsect-Liu}
Assuming $\tom$ and $\tq$ are of the same order, the solution of Eq. \eqref{eomGBsm} was found in Appendix B1 of Ref. \cite{Brigante:2007nu}. It is of the form \eqref{decZsm} with $g(\fz)$ given by Eq. (B3) therein,
\begin{equation}\label{gLiu}
g(\fz)=1+\frac{i\tq}{4W}\left(1-\frac{1}{\fz^4}\right)\left[1+\lagb\left(3(W^2-1)-\frac{1}{\fz^4}\right)\right]+\O(\tq^2,\lagb^2),
\end{equation}
where $W=\tom/(\tq\Nsh)$ is a parameter introduced in Refs. \cite{Kovtun:2005ev,Brigante:2007nu}. Then the damping constant \eqref{DamGBsm} was verified under the assumption $\tom\sim\O(\tq^2)$ in Ref. \cite{Brigante:2007nu}. In the present subsection, we will elucidate that, to get the above solution, there was an implicit requirement $g_1(\fz)=0$ with $g_1(\fz)$ defined in Eq. \eqref{decZLiu}.

According to the assumption $\tom\sim\O(\tq)$, after rescaling $\tom\rightarrow\lambda\tom$, $\tq\rightarrow\lambda\tq$, the solution \eqref{decZsm} can be expanded in powers of $\lambda$ as
\begin{eqnarray}\label{decZLiu}
\nonumber Z(\fz,\lambda\tom,\lambda\tq)&\approx&\left(1-\frac{1}{\fz^4}\right)^{-i\lambda\tom/(4\Nsh)}\left[g_0(\fz)+\lambda g_1(\fz)+\O(\lambda^2)\right]\\
&&\approx g_0+\lambda\left[g_1-\left(\frac{i\tom}{4\Nsh}\ln\frac{\fz^4-1}{\fz^4}\right)g_0\right]+\O(\lambda^2)
\end{eqnarray}
in which $g_0(\fz)$, $g_1(\fz)$ are regular at $\fz=1$. The equation of motion \eqref{eomGBsm} can be expanded as
\begin{equation}\label{eomLiu}
\partial_{\fz}^2Z+\left[\frac{5\fz^4-1}{\fz^4-1}+\frac{4}{(W^2-1)\fz^4+1}\right]\frac{1}{\fz}\partial_{\fz}Z-\frac{4\lagb\left[2(\fz^4-1)^2+4W^2\fz^4-3W^4\fz^8\right]}{\fz^5\left[(W^2-1)\fz^4+1\right]^2}\partial_{\fz}Z+\O(\lambda^2)\approx0.
\end{equation}
From this equation, we can see that $g_0$ and $g_1-\left(\frac{i\tom}{4\Nsh}\ln\frac{\fz^4-1}{\fz^4}\right)g_0$ obey the same equation, thus they should have the same general solution. This is in contrast to other sections, where the $\O(1)$ equations of motion are different from the $\O(\lambda)$ ones and less complicated than Eq. \eqref{eomLiu}. In order to find the general solution of Eq. \eqref{eomLiu}, we rewrite this equation as
\begin{equation}
\partial_{\fz}\left\{\frac{\fz^5(\fz^4-1)}{(W^2-1)\fz^4+1}\exp\left[\frac{2\lagb}{\fz^4}-\frac{\lagb W^2(5W^2-4)}{(W^2-1)^2\fz^4+W^2-1}\right]\partial_{\fz}Z\right\}+\O(\lambda^2)\approx0.
\end{equation}
Neglecting $\O(\lagb^2)$ terms, its solution has the nice analytic form
\begin{eqnarray}\label{ZLiu}
\nonumber Z(\fz,\lambda\tom,\lambda\tq)&\approx&-\frac{(\lambda C_3+C_1)W^2}{4}\left[1+\frac{\lagb(3W^2-2)}{W^2-1}\right]\ln\frac{\fz^4-1}{\fz^4}\\
&&-\frac{\lambda C_3+C_1}{4\fz^4}\left[1+\frac{\lagb W^2(3W^2-2)}{W^2-1}-\frac{\lagb}{\fz^4}\right]+\lambda C_2+C_0+\O(\lambda^2)
\end{eqnarray}
Matching this solution to Eq. \eqref{decZLiu} gives us
\begin{eqnarray}
\label{C1Liu}C_1&=&0,\\
\label{C3Liu}C_3&=&\frac{iC_0\tom}{\Nsh W^2}\left[1+\frac{\lagb(3W^2-2)}{W^2-1}\right]^{-1},
\end{eqnarray}
\begin{eqnarray}
g_0&=&C_0,\label{g0Liu}\\
g_1&\approx&-\frac{iC_0\tq}{4W\fz^4}\left[1+\lagb(3W^2-2)-\frac{\lagb}{\fz^4}\right]+C_2,\label{g1Liu}
\end{eqnarray}

Confronting Eq. \eqref{gLiu} with Eqs. \eqref{g0Liu}, \eqref{g1Liu}, one can see
\begin{equation}
C_2=\frac{iC_0\tq}{4W}\left[1+3\lagb(W^2-1)\right]
\end{equation}
or equivalently $g_1(1)=0$. Conversely, to get Eq. \eqref{gLiu} from Eqs. \eqref{g0Liu}, \eqref{g1Liu}, one has to require that $g_1(1)=0$.

In Appendix C of Ref. \cite{Grozdanov:2016fkt}, the boundary condition $g_0(1)=0$, $g_i(1)=0$ for $i>0$ was justified by the fact that in holography we work with bulk solutions normalized to one at the boundary. Interestingly, our calculations in Secs. \ref{sect-Sch} and \ref{sect-GBsm} required only that $g_0(1)$ and $g_i(1)$ not be divergent.

Taking $\lagb\rightarrow0$, the discussion in this subsection has some relevance also to Sec. IVB2 of Ref. \cite{Kovtun:2005ev} regarding the shear channel of Schwarzschild black brane.

\section{Discussion}\label{sect-disc}
In our notes, we have refined the recipe for computing hydrodynamic dispersion relation of quasinormal modes, and have demonstrated it in the diffusive channel of the Schwarzschild black brane and the shear channel of the Gauss-Bonnet black brane. All through the refined recipe, we got rid of the assumption $\omega\sim\O(q)$ and thus worked with $\omega\sim\O(q^2)$ in the leading order. We have also elucidated that, in the literature, an implicit requirement was imposed on the $\O(\omega,q)$ wave functions. In the refined recipe, there is no such requirement.

Although some technical details have been refined in our notes, the main scheme and final results of Refs. \cite{Kovtun:2005ev,Starinets:2008fb,Brigante:2007nu} are never ruined. In fact, our work reinforces their scheme and confirmes their results. After the refinement, the recipe becomes less confusing.

We have restricted our investigation to the $\O(q^2)$ term in Eq. \eqref{dispexp}. For the Gauss-Bonnet black brane, we have focused on the small $\lagb$ case and neglected $\O(\lagb^2)$ terms. This helps to make the notes transparent. However, the recipe can be readily applied to the Gauss-Bonnet black brane with arbitrary $\lagb$ and to the $\O(q^4)$ term or higher order terms. To avoid distraction, we report our results in an accompanying paper \cite{Wang:2019xny}.

\begin{acknowledgments}
This work is supported by the National Natural Science Foundation of China (Grant No. 91536218). The authors would like to thank the anonymous referee for valuable comments on an earlier version of this paper.
\end{acknowledgments}

\appendix

\section{More on the $\O(q)$ term in $\omega$}\label{app-lin}
In Sec. \ref{sect-lin}, we have taken a closer look at the customary recipe which assumes firstly $\fw\sim\O(\fq)$ and finally $\fw\sim\O(\fq^2)$. What can we get if we adhere to the assumption $\fw\sim\O(\fq)$? We will explore this question in the current appendix.

For the Schwarzschild black brane in the Einstein gravity, the solution of perturbations in the diffusive channel is given by Eq. \eqref{ExStar}, in which we have set the parameter $\lambda=1$. Recovering $\lambda$ to highlight the orders of the perturbation, we find it takes the form
\begin{equation}
E_x(u,\lambda\fw,\lambda\fq)=C_0-\frac{i\lambda}{2\fw}f'(1)C_0\sqrt{-g}g^{tt}g^{uu}\int_0^u\frac{\fw^2-\fq^2f}{f\sqrt{-g}g^{tt}g^{uu}}du+\lambda C_2+\O(\lambda^2).
\end{equation}
At spatial infinity, $u=0$, it becomes
\begin{equation}
E_x(0,\lambda\fw,\lambda\fq^2)=C_0+\lambda C_2+\O(\lambda^2).
\end{equation}
According to traditional perturbation theory, similar to what we have done in Sec. \ref{sect-Sch}, the boundary condition $E_x(0)=0$ tells us that $C_0=C_2=0$. In other words,
\begin{equation}
E_x(u,\lambda\fw,\lambda\fq)=0+\O(\lambda^2).
\end{equation}

For the black brane in the Gauss-Bonnet gravity, the solution of perturbations in the shear channel can be studied in the same way. At spatial infinity, $\fz\rightarrow\infty$, imposing the boundary condition $Z(\infty)=0$ on Eq. \eqref{ZLiu} yields $C_0=C_2=0$ according to traditional perturbation theory. In combination with Eqs.\eqref{C1Liu}, \eqref{C3Liu}, it leads to
\begin{equation}
Z(\fz,\lambda\tom,\lambda\tq)=0+\O(\lambda^2)+\O(\lagb^2).
\end{equation}

These results suggest that the equations of motion with the Dirichlet boundary condition $E_x=0$ or $Z=0$ at spatial infinity have no solution compatible with the assumption $\fw\sim\O(\fq)\ll1$ up to $\O(\fq)$. This motivates the dispersion relation \eqref{dispexp} starting with the $q^2$-term in the low-frequency limit.

\end{document}